# Root Distribution in Padé Approximants and its Effect on Holomorphic Embedding Method Convergence

Songyan Li, *Student Member, IEEE,* Abhinav Dronamraju, *Student Member, IEEE*, Daniel Tylavsky, *Life Senior Member, IEEE*

*Abstract*—The requirement for solving nonlinear algebraic equations is ubiquitous in the field of electric power system simulations. While Newton-based methods have been used to advantage, they sometimes do not converge, leaving the user wondering whether a solution exists. In addition to improved robustness, one advantage of holomorphic embedding methods (HEM) is that, even when they do not converge, roots plots of the Padé approximants (PAs) to the functions in the inverse-α plane can be used to determine whether a solution exist. The convergence factor (CF) of the near-diagonal PAs applied to functions expanded about the origin is determined by the logarithmic capacity of the associated branch cut (BC) and the distance of the evaluation point from the origin. However the underlying mechanism governing this rate has been obscure. We prove that the ultimate distribution of the PA roots on the BC in the complex plane converges weakly to the equilibrium distribution of electrostatic charges on a 2-D conductor system with the same topology in a physical setting. This, along with properties of the Maclaurin series can be used to explain the structure of the CF equation We demonstrate the theoretical convergence behavior, with numerical experiments.

## I. Introduction

The robust solvers for nonlinear power system problems currently used in industry typically consist of using several different algorithms in the hopes that, for any ill-condition problem, one of them will converge to a solution. These solvers use proprietary heuristics that have been found on average to yield better results. And yet, not all problems with feasible solutions converge to an operable solution. The previous statement is true regardless of the nonlinear problem of interest: power flow, state estimation, optimal power flow, etc.

Since its introduction to the power-system community in 2012, the holomorphic embedding method (HEM), [1], [2], has attracted a great deal of attention. In addition to being extensively studied as a solution to the power-flow problem in general, and the ill-conditioned problem specifically, [3]-[7], it has been applied to unit commitment, [8], reduced-order network equivalents, [9]-[11], special devices, [12], [13], and voltage-stability-margin estimation, [14]-[20]. HEM has been expanded to handle multi-dimensional embeddings [21]-[24], applied to dynamic simulations, [25], used to attack the multivalued problem [26]-[29] and enhanced to improve robustness and computational efficiency, [30]-[36]. One of the reasons for the broad interest in HEM, was the claim of a universal convergence guarantee [1], (provided a solution existed) ostensibly supported by Stahl's theorems, [46]-[51]. The scope of that putative universal guarantee was broad, apparently independent of network parameters, and encompassing theoretical as well as numerical convergence.

Eventually, limitations to the theoretical and numerical convergence guarantees suggested by [3] and [37] indicated that, using the current embeddings and existing theory, neither theoretical nor numerical convergence was guaranteed. This controversy led to investigations that laid out the requirements for theoretical and numerical convergence [39], [40].

For numerical convergence to be attained, the convergence factor (CF) (which is a function of the branch-cut capacity (BCC) and distance from the point of expansion) must be sufficiently small, so that convergence to the required tolerance is achieved within the number of recursion terms dictated by the precision of the floating-point numbers used [40]. To guarantee theoretical convergence to an operable solution (provided one exists), it is sufficient to prove that all of the branch points of the functions lie inside a circle centered at the origin (of the so-called inverse-α plane) with a radius less than the point of interest. One attempt to prove theoretical convergence for a particular embedding has met with some success by providing local convergence guarantees for the embedding studied [38].

In some ways, we are now at the point with HEM that we were in the 1950's and early 1960's with Newton's method, which was first proposed in 1956 [41]. Until Sato and Tinney suggested sparsity programming techniques [42] in 1963 and Tinney and Walker formalized optimal ordering in 1967 [43], Newton's method, which had been languishing for 11 years, was largely a university research curiosity and/or relegated to small system problems in industry. Even after these innovations, years would elapse before Newton-based methods found wide acceptance. Now it, and its many variants, are the workhorses for solving sets of nonlinear algebraic equations for power system problems.

Newton's method, and other iteration methods, characterized as 'displacements of the center,' have convergence behavior that changes with each new 'center' point or, point of development. The iterates can lead to places where the linear approximation (slope) of function is a poor approximation to the function as evidenced by unpredictable jumps in the mismatch-versus-iteration-number plots, making convergence behavior hard to model and predict. In contrast, the point of development of a HEM approach, based on the Maclaurin series of the function, remains at the origin, allowing the behavior of the convergence factor (CF) (inverse of the convergence rate), to be given by,

$$CF(\alpha) = |\alpha|BCC + O(\alpha^2) \text{ as } \alpha \to 0 \qquad (1)$$

where α is the embedding parameter and $O(\alpha^2)$ represents higher order terms [40].

The goal of this paper is twofold: to uncover the origins of the CF behavior as it grows out of both the roots behavior of the

Songyan Li, Abhinav Dronamraju and Daniel Tylavsky are with the School of Electrical Computer and Energy Engineering in Arizona State University, Tempe, AZ 85287, USA (email: songyanl@asu.edu / dronamraju@asu.edu / tylavsky@asu.edu).

sequence of PAs and properties of the Maclaurin series.

We begin by observing that the roots of a near-diagonal Padé approximant (PA) corresponding to the Maclaurin series generated by HEM accumulate on the branch cut (BC) with minimum logarithmic capacity. The convergence behavior is determined by the number and placement of these roots. This provides a starting point for understanding the convergence behavior as suggested by [39]: "…the (ultimate) root distribution on the branch cut defined by the PA is identical to (a discretized) electrostatic charge distribution at equilibrium on a conductor with the two-dimensional topology of the branch cut as it exists in the inverse-α plane." Here, a "two-dimensional" (2-D) topology is shorthand for a three-dimensional conductor system that is homogeneous in one of the orthogonal directions.

Establishing that the near-diagonal PA root distribution may be predicted using an equivalent electrostatic problem as suggested [39] is the first step in understanding the origin of (1) and is the main contribution of this paper.

While the equivalence of the PA root distribution to the equilibrium distribution in an electrostatic system was claimed by [44] to be proven in [45], it is clear this is not case and [45] makes no claim to such a proof. The proof of this equivalence is established in this work and the implications of that result on the convergence behavior are developed with numerical examples.

Finally, the proof contained here relies on the concept of logarithmic capacity. To be consistent with the definitions used by Stahl in his proofs [46], we shall be using the consistent definition of logarithmic capacity in [48] and [53], not the one found in[54].

## II. Précis

We prove the following theorem and describe its implications to PA convergence in a HEM-based algorithm, using the PF problem as an exemplar.

*Theorem. 1: Let the normalized Borel measure $\mu(\mathbf{r})$ be the per-unit-length electrostatic charge measure describing the charge distributed on a three-dimensional foil conductor of topology E which is homogeneous along one orthogonal coordinate. Define the electrostatic energy integral as,*

$$K[\mu] = \int_E \int_E \ln \frac{1}{|\mathbf{r} - \mathbf{r}'|} d\mu(\mathbf{r}') d\mu(\mathbf{r}) \quad (2)$$

*where r is a point on E, then the charge measure that is consistent with charge distributed on a 2-D capacitor with topology E, defined by,*

$$W(E) = \inf_\mu K[\mu] \quad (3)$$

*is identical to the normalized equilibrium measure in potential theory for a branch cut in the complex plane of the same topology.*

One implication of Thm. 1 is that the support of $\mu$, E, is, for the present purposes, the conducting foil upon which the charge is distributed and that no charge is distributed in the free space. Compare this theorem with the following theorem defining logarithmic capacity, $cap(E)$, as it applies to defining the BC for a PA.

*Theorem 6.6.6 [48]. Let $\mu(x)$ be a normalized measure defined on E and define,*

$$I[\mu] = \int_E \int_E \ln \frac{1}{|z - z'|} d\mu(z) d\mu(z') \quad (4)$$

*Then let,*

$$V(E) = \inf_\mu I[\mu] \quad (5)$$

*Then*

$$cap(E) = e^{-V(E)} \quad (6)$$

Finding the BC with minimum logarithmic capacity, $cap(E)$, which is the BC toward which the roots of the PA accumulate, requires minimizing (6) over all possible topologies, E, where (6) provides the definition of logarithmic capacity.

Observe that the integrals in (2) and (4) are essentially identical; the integral in (2) is over the (real) 2-D x-y plane and the integral in (4) is over the 2-D complex plane. Observe that the minimization over all $\mu$ in (5), is identical to the minimization of $K[\mu]$ required by Thomson's theorem, [59], [60], in the electrostatic case, which is discussed below.

In short, given a conducting foil in the x-y plane with the same topology, E, as the BC in the complex plane, the electrostatic charge measure, $\mu$, consistent with (2) and (3) must be identical to the unique equilibrium measure, $\mu$, defined by (4) and (5). Said another way, as the number of terms in the PA grows, the root distribution of the PAs more closely approximates the equilibrium electrostatic charge distribution consistent with the electrostatic charge measure. This suggests the rate of change of the difference between the equilibrium electrostatic charge distribution and the PA root distribution, informs the convergence behavior defined by (1).

In the next section, we discuss the meaning of (4)-(6) in terms of measure theory. In the subsequent section, the development of (2) from first principles is reported. The sections after that provide numerical evidence confirming the connection between these equilibrium measures.

## III. The Logarithmic Energy and the Logarithmic Capacity

The terminology used to discuss to Thm. 6.6.6, which is a result that comes from potential theory, is likely new to many engineers and we use this section to level-set the discussion. First, the double integral defined in (4) uses measure theory and is called the logarithmic energy [53], where $\mu$ is the underlying (normalized) Borel measure [53], [55], [56]. The measure, for which the infimum in (5) is attained, is unique and is called the equilibrium measure in potential theory [53]. In real analysis, an integral, like (4), involving a measure is defined using the so-called simple functions and a limit procedure in a rigorous and profound way [55], which is outside the scope of this paper. The symbol $d\mu(z)$ in (4) is equivalent to $\mu(dz)$ [57], which means, for our purposes, $d\mu(z)$ is the value of $\mu$ when it acts on $dz$, where $dz$ could be regarded as the generalized infinitesimal volume (area/length) (the word "generalized" is subsequently suppressed) in a 3-D (2-D/1-D) space. In the case of PAs, $\mu$ represents a measure of root density. If $\mu$ is assigned a physical meaning, for example, charge in an electrostatic system, then $\mu(dz)$ is (a measure of) the total charge contained in the infinitesimal space $dz$. A normalized Borel measure, $\mu(z)$, on a set E is defined as:

$$\int_E d\mu(z) = \mu(E) = \text{a fixed constant} \quad (7)$$
$$= 1|_{for\ equlibrium\ measure}$$

If $\mu$ is an abstract measure of charge then (7) defines the total charge contained in the set (space) $E$ as a constant. This constant must be defined to be 1 (appendix of [53]), i.e. $\mu(E) = 1$, to obtain the equilibrium measure.

It has been shown that the normalized measure counting the poles and the zeros of the sequence of near-diagonal PAs (considering multiplicity) converges weakly to the equilibrium measure on the set where the PAs do not converge, which is the BC with minimum logarithmic capacity (Thm. 1.8 in [46]). (In the context of measure and potential theory, strong convergence implies the measure converges to the equilibrium measure $\mu_n \to \mu$. Weak convergence implies only that the sequence of expected values converges to the expected value of the equilibrium measure, $\int f d\mu_n \to \int f d\mu$, for any bounded measurable $f$.) If $\mu$ is a measure of the number of roots of the near-diagonal PA on the BC, this suggests that there may always be some roots not on the BC, as hypothesized by Nuttal [51]. Observe if $\mu$ converged strongly, the PAs for all bus voltages in the multi-equation power flow problem, which share the same BC, would have identical roots as the number of terms in the defining Maclaurin series grew without bound, resulting in the PV curve of all buses having the identical shape.

## IV. THE CHARGE DISTRIBUTION OF A 2-D ELECTROSTATIC SYSTEM WITH 2-D CONDUCTORS

In this section we show that the equilibrium measure defined in the previous section is the mathematical abstraction/analog of the physical electrostatic equilibrium charge distributed on a 2-D conductor topology which is identical to the BC.

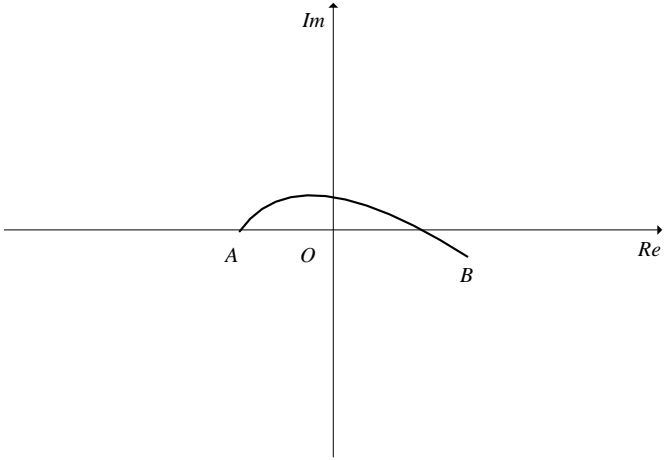

Fig. 1 One possible curve in the complex plane

### A. The curve and the foil correspondence

Let's assumed that the BC where the PAs do not converge is the Jordan arc in the complex plane connecting points $A$ and $B$ in Fig. 1, where the horizontal axis is the real axis and the vertical axis is the imaginary axis.

How does the curve shown in Fig. 1 correspond to an object in the 3-D space? If the real (imaginary) axis is assigned to the x-axis (y-axis), the z-axis can be added orthogonally to the x-y plane, forming a right-handed coordinate system. Translations of the curve along the z-axis to $\pm\infty$ makes a surface, along the z-axis, which looks like a finitely wide but infinitely long foil running parallel to the z-axis, shown in Fig. 2. Because this surface is homogeneous about the z-axis, 2-D modeling is sufficient to describe the electric field and this foil is referred to as a 2-D conductor.

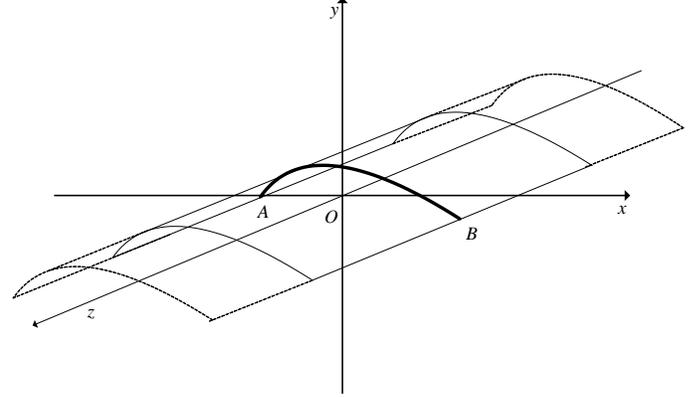

Fig. 2 The 3-D shape of the 2-D set

### B. The electrostatic energy of a 3-D system homogeneous in one direction

Next, we will show that the equilibrium measure in potential theory is just the mathematical abstraction of the charge distributed on the foil shown in Fig. 2 if the foil is treated as ideal conductor. The Poisson equation in electrostatics is well known as shown in (8),

$$-\nabla^2 u = -\left(\frac{\partial^2 u}{\partial x^2} + \frac{\partial^2 u}{\partial y^2} + \frac{\partial^2 u}{\partial z^2}\right) = \frac{\rho(r)}{\varepsilon_0} \quad (8)$$

where $\rho(r)$ is the space charge density, $r$ is the position vector $(x, y, z)$, $\varepsilon_0$ is the vacuum permittivity and the function that satisfies this equation, $u$, is the electrostatic potential. Assuming that the system is homogeneous along the z-axis direction, $\rho(x, y, z) = \rho(x, y)$ with its units, C/m³, unchanged, and the partial derivative with respect to the z-coordinate vanishes, giving us,

$$-\left(\frac{\partial^2}{\partial x^2} + \frac{\partial^2}{\partial y^2}\right)u = \frac{\rho}{\varepsilon_0} \quad (9)$$

which is the 2-D Poisson equation. The position vector $r$ becomes $(x, y)$, and the solution of Poisson's equation, i.e., the electrostatic potential, does not depend on the z-coordinate.

When charge is distributed on a surface with zero thickness, as is the case with our 2-D conductor foil, the space charge density on this surface cannot be well-defined to be finite in general, but the potential clearly exists in the free space. In such cases, one can use a charge measure ($\mu$), bypassing defining $\rho$, as in (9). Our immediate goal is to rigorously show that using a charge measure is theoretically appropriate.

The function shown in (10) is proposed as the electric potential of a 3-D electrostatic system that is homogeneous along the z-axis direction,

$$u(r) = \frac{1}{2\pi\varepsilon_0}\int \ln\frac{a}{|r - r'|}d\mu(r') \quad (10)$$

where $a$ is an arbitrary constant with units of distance and $\mu(r')$, has units of charge per-unit-length along the z direction. (The

integration domain, $E$, is suppressed in this and the following integral equations.)

To see why (10) is consistent with an interpretation as an electrostatic potential, we show both that it obeys Poisson's equation when the Radon–Nikodym density (the math abstraction of the charge density) is well-defined and that $\mu(A)$ is consistent with the definition of per-unit-length charge contained in the cross-section area $A$.

Following the approach used in [56], $\mu$ is assumed to be absolutely continuous, [55], which means the Radon–Nikodym density corresponding to $\mu$ is well-defined. Then (10) can be written as,

$$u(\boldsymbol{r}) = \frac{1}{2\pi\varepsilon_0} \int \ln\frac{a}{|\boldsymbol{r}-\boldsymbol{r}'|} \hat{\rho}(\boldsymbol{r}') d^2\boldsymbol{r}' \tag{11}$$

where $\hat{\rho}(\boldsymbol{r}')$ is the corresponding Radon–Nikodym density [55], [56], and $d^2\boldsymbol{r}'$ is the infinitesimal area element, usually written as $d\sigma'$ or $dx'dy'$. (Absolutely continuous essentially means that the electrostatic charge density is well-defined and $d\mu(\boldsymbol{r}') = \rho(\boldsymbol{r}')d^2\boldsymbol{r}'$, is the per-unit-length (along the z-axis direction) charge stored in the tube whose cross-section area is $d^2\boldsymbol{r}'$.) It is well known that,

$$-\left(\frac{\partial^2}{\partial x^2} + \frac{\partial^2}{\partial y^2}\right)\left(\frac{1}{2\pi}\ln\frac{1}{|\boldsymbol{r}-\boldsymbol{r}'|}\right) = \delta(\boldsymbol{r}-\boldsymbol{r}') \tag{12}$$

where $\delta(\boldsymbol{r}-\boldsymbol{r}')$ is the 2-D Dirac delta function [58]. After substituting (11) back into the LHS of (9), and exchanging the order of the 2-D Laplacian and the integral, one gets,

$$\begin{aligned}
&-\left(\frac{\partial^2}{\partial x^2} + \frac{\partial^2}{\partial y^2}\right)u(\boldsymbol{r}) \\
&= \frac{1}{\varepsilon_0}\int -\left(\frac{\partial^2}{\partial x^2} + \frac{\partial^2}{\partial y^2}\right)\left(\frac{1}{2\pi}\ln\frac{1}{|\boldsymbol{r}-\boldsymbol{r}'|} + \frac{\ln a}{2\pi}\right)\hat{\rho}(\boldsymbol{r}')d^2\boldsymbol{r}' \\
&= \frac{1}{\varepsilon_0}\int -\left(\frac{\partial^2}{\partial x^2} + \frac{\partial^2}{\partial y^2}\right)\left(\frac{1}{2\pi}\ln\frac{1}{|\boldsymbol{r}-\boldsymbol{r}'|}\right)\hat{\rho}(\boldsymbol{r}')d^2\boldsymbol{r}' \\
&= \frac{1}{\varepsilon_0}\int \delta(\boldsymbol{r}-\boldsymbol{r}')\hat{\rho}(\boldsymbol{r}')d^2\boldsymbol{r}' = \frac{\hat{\rho}(\boldsymbol{r})}{\varepsilon_0}
\end{aligned} \tag{13}$$

which proves that (11) satisfies the Poisson equation and it has the physical meaning of an electrostatic potential, where the Radon–Nikodym density $\hat{\rho}(\boldsymbol{r}')$ is the abstraction of the physical space charge density in (8). Therefore $\hat{\rho}(\boldsymbol{r}') = \rho(\boldsymbol{r}')$ if they are both well-defined.

It can be seen that total charge, stored in a slice of space whose thickness is $\Delta L$ along the z-axis direction, shown in Fig. 3, can be calculated by,

$$\begin{aligned}
\Delta Q &= \int \rho(\boldsymbol{r})d^3\boldsymbol{r} = \int_0^{\Delta L} dz \int \rho(\boldsymbol{r})d^2\boldsymbol{r} \\
&= \Delta L \int \rho(\boldsymbol{r})d^2\boldsymbol{r} = \Delta L \int d\mu(\boldsymbol{r}) = \int \Delta L d\mu(\boldsymbol{r})
\end{aligned} \tag{14}$$

where $d^3\boldsymbol{r}$ is the infinitesimal volume element usually written as $dV$ or $dxdydz$.

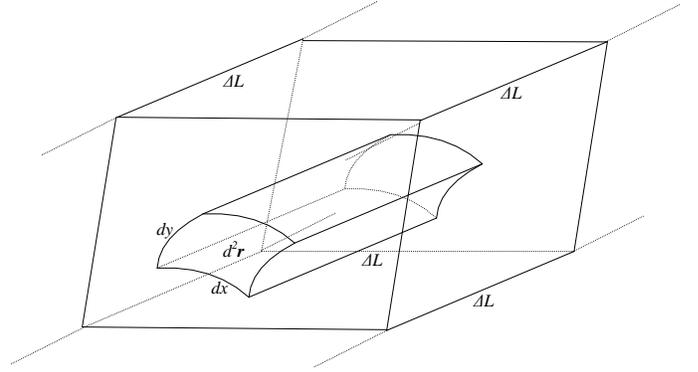

Fig. 3 The slice of space with finite thickness

With $d\mu(\boldsymbol{r})$ as the per-unit-length charge stored in the tube whose cross-section area is $d^2\boldsymbol{r}$, at position $\boldsymbol{r}$, $\Delta L\, d\mu(\boldsymbol{r})$ is the total charge stored in this tube whose length is $\Delta L$. The final integral in (14) simply says that by summing up all the charges stored in the tubes that form a slice of space whose thickness is $\Delta L$, one gets the total charge stored in that slice of space. Then it is obvious that for the cross-section area $A$, the per-unit-length charge stored is,

$$\left.\frac{\Delta Q}{\Delta L}\right|_A = \int_A d\mu(\boldsymbol{r}) = \mu(A) \tag{15}$$

We assume that a charge density is well defined when defining (11) as a potential, but the above work shows that this assumption of a formulation involving a space charge density can be bypassed when using the measure $\mu$ with $d\mu(\boldsymbol{r})$ as the per-unit-length charge stored in the tube whose cross-section area is $d^2\boldsymbol{r}$ at position $\boldsymbol{r}$. With this in mind, the total electrostatic energy stored in the slice of space whose thickness is $\Delta L$ is calculated next.

Given the electrostatic potential written in (10), the equation for calculating the electrostatic energy [59] is,

$$E_e = \frac{1}{2}\int u(\boldsymbol{r})\rho(\boldsymbol{r})d^3\boldsymbol{r} \tag{16}$$

Based on this equation, the total electrostatic energy stored in the slice of space whose thickness is $\Delta L$ along the z-axis direction is (remembering that $d\mu(\boldsymbol{r}) = \rho(\boldsymbol{r})d^2\boldsymbol{r}$ if $\rho(\boldsymbol{r})$ is well defined) with its physical meaning shown by (14) and (15)),

$$\begin{aligned}
\Delta E &= \frac{1}{2}\int u(\boldsymbol{r})\rho(\boldsymbol{r})d^3\boldsymbol{r} \\
&= \frac{1}{2}\int_0^{\Delta L} dz \int u(\boldsymbol{r})\rho(\boldsymbol{r})d^2\boldsymbol{r} \\
&= \frac{\Delta L}{2}\int u(\boldsymbol{r})\rho(\boldsymbol{r})d^2\boldsymbol{r} = \frac{\Delta L}{2}\int u(\boldsymbol{r})d\mu(\boldsymbol{r}) \\
&= \frac{\Delta L}{4\pi\varepsilon_0}\iint \ln\frac{a}{|\boldsymbol{r}-\boldsymbol{r}'|} d\mu(\boldsymbol{r}')\, d\mu(\boldsymbol{r})
\end{aligned} \tag{17}$$

Using (16), the per-unit-length electrostatic energy is,

$$\frac{\Delta E}{\Delta L} = \frac{1}{4\pi\varepsilon_0}\iint_E \ln\frac{a}{|\boldsymbol{r}-\boldsymbol{r}'|} d\mu(\boldsymbol{r}')d\mu(\boldsymbol{r}) \tag{18}$$

and one immediately recognizes that with the exception of a positive constant factor and an arbitrary distance constant, $a$, the per-length electrostatic energy's functional form, (18), is the same as the logarithmic energy, (4). While the integration domain $E$ represents the locations where the charge resides in (18), the set $E$ in (4) represents the BC topology.

*C. The equilibrium measure and the electrostatic charge distributed on a conductor correspondence*

Thomson's theorem states that the electric charge on a set of conductors distributes itself on the conductor surfaces to minimize the electrostatic energy, [59], [60]. This means that when the foil shown in Fig. 2 is considered as a conductor, the charge distributed on it must minimize the per-unit length electrostatic energy shown in (18), assuming that all the charge is distributed on the conductor and there is no charge in the free space. To see how the equilibrium measure is related to the equilibrium charge as it is distributed on the conductor, (18) is further developed as,

$$\frac{\Delta E}{\Delta L} = \frac{1}{4\pi\varepsilon_0} \int \int \ln\frac{1}{|\boldsymbol{r}-\boldsymbol{r}'|} d\mu(\boldsymbol{r}') \, d\mu(\boldsymbol{r})$$
$$+ \frac{\ln a}{4\pi\varepsilon_0} \int \int d\mu(\boldsymbol{r}') \, d\mu(\boldsymbol{r}) \quad (19)$$

and referring to (14), one gets,

$$\frac{\Delta E}{\Delta L} = \frac{1}{4\pi\varepsilon_0} \int \int \ln\frac{1}{|\boldsymbol{r}-\boldsymbol{r}'|} d\mu(\boldsymbol{r}') \, d\mu(\boldsymbol{r})$$
$$+ \frac{\ln a}{4\pi\varepsilon_0} \left(\frac{\Delta Q}{\Delta L}\right)^2 \quad (20)$$

If the charge is normalized by setting the per-unit-length charge to a fixed constant, e.g., 1,

$$Q_{pL} = \frac{\Delta Q}{\Delta L} = \int d\mu(\boldsymbol{r}) = 1 \quad (21)$$

the per-length electrostatic energy becomes,

$$E_{pL} = \frac{\Delta E}{\Delta L}$$
$$= \frac{1}{4\pi\varepsilon_0} \int \int \ln\frac{1}{|\boldsymbol{r}-\boldsymbol{r}'|} d\mu(\boldsymbol{r}') \, d\mu(\boldsymbol{r}) + \frac{\ln a}{4\pi\varepsilon_0} \quad (22)$$

Because the second term is a constant and the factor of the first term is positive and with the normalization (21), minimizing the per-length electrostatic energy is the same as minimizing,

$$\int \int \ln\frac{1}{|\boldsymbol{r}-\boldsymbol{r}'|} d\mu(\boldsymbol{r}') \, d\mu(\boldsymbol{r}) \quad (23)$$

which is mathematically the same as minimizing the logarithmic energy in (4). From the definition of the equilibrium measure, it is immediately obvious that equilibrium measure is, once again, the mathematical abstraction of the charge distributed on a conductor that is homogeneous along the z-axis direction.

The branch cut on the complex plane may have disconnected curves and their counterparts in the 3-D space are graphically disconnected conductor foils (homogeneous along the z-axis direction) but these foils are electrically connected, because the normalization condition, (21), requires that the total per-unit-length charge is fixed and individual charges can be placed on any foil freely as long as (21) is satisfied and charges can be move freely from one foil to another to minimize the per-length energy.

Theorem 1.8 in [46] states that the normalized measures that count the poles and zeros of the near diagonal Padé approximants weakly converge to the equilibrium measure. By the derivation shown in this section, we conclude that those normalized measures converge weakly to the normalized charge distributed on the conductor foils, homogeneous in one direction, whose cross sections are the same as the branch cuts. This explains why the poles and zeros of the Padé approximants accumulates more near the tips of the branch cut: because in electrostatics, the charges gravitate toward the ends of conductors because of electrostatic repulsion.

*D. The logarithmic capacity and the electrostatic energy correspondence*

Using the definition of the logarithmic capacity, (4), (5) and (6), and the per-unit-length electrostatic energy, (22), and assuming the arbitrary distance constant is 1, we obtain the expression, (24), relating the logarithmic capacity to the per-unit-length electrostatic energy of the conducting foil system whose cross section is the same as the branch cut.

$$C_E = cap(E) = e^{-4\pi\varepsilon_0 E_{pL}} \quad (24)$$

If the BC is a straight-line segment, according to [48], the longer it is, the larger the logarithmic capacity, and the smaller the stored energy according to (6). For a 3-D finite conductor system, where the charge is normalized to a fixed constant, say 1, from the equation for electrical energy, $E_e$, stored in an electrical capacitor, the relationship between capacitance and energy is,

$$C_e = \frac{Q^2}{2E_e} = \frac{1}{2E_e} \quad (25)$$

where the monotonically decreasing value of logarithmic capacity/electrostatic capacitance with energy in (24) and (25), respectively, for a fixed amount of charge, is maintained, as one would hope. That the logarithmic capacity and electrostatic capacitance behave similarly with energy provides a conceptual framework for developing an intuitive understand of the behavior of BCC on convergence.

*E. Quantization*

We have shown how the continuous equilibrium electrostatic charge distribution of a 2-D capacitor is connected to the root distribution (of the near-diagonal) PA and consistent with the equilibrium measure of a BC, given identical topologies in both cases. When a finite number of terms is use in the PA, each root corresponds to one quantum of charge in the analogous electrostatic problem, with the total charge, equal to that of the equilibrium Borel measure in the logarithmic energy problem. As the number of PA roots increases, the discretized root distribution converges weakly to the continuous electrostatic distribution.

V. NUMERICAL RESULTS

Using a two-bus system, where the developed theory fits the numerical root placement and convergence behavior plots in a visually obvious way, allows the connection of the theory and numerics to be seen most clearly. Using that example as a jumping-off point, we will see the same pattern with the higher-order 118-bus system, with a more complex BC.

*A. Two-Bus Asymmetrical System CF Behavior*

We numerically verify the electrostatic-charge/PA-root-distribution 'equivalence relationship' using a two-bus system power-flow problem with a BC of (logarithmic) capacity 1, colinear with the real axis in the complex plane. The classical form HEM embedding [39] was used, with expansion about the zero in the α-plane (αP). The loading was selected to create a BC that is asymmetric about the origin with branch points in the inverse α plane (IαP) at -3 and +1 (-1/3 and +1 in the αP). We refer to this as the Asymmetrical System (AS).

The convergence factor (CF) plotted versus α in Fig. 4 was calculated using numerical voltage-error results and conforms with (1) asymptotically near the point of development, the origin, where the slope of the curve is approximately ± 1.0, the magnitude of the BCC. Also as expected, CF approaches 1.0 as α approaches the branch points, beyond which no solution exists in the field of complex numbers. Even though theoretical convergence is possible as one approaches a branch point, finite numerical precision, which limits the number of terms effective in constructing the PA, leads to numerical nonconvergence as one reaches this boundary in the convergence domain [40].

With these results, we can explain the CF dependence on α in (1): the CF behavior vs. α in Fig. 4 (for high degree PAs) is inherited from its Maclaurin series input given 1) that the PA may be viewed as a generalization of the Maclaurin series to the class of rational functions [52] and, 2) that the Maclaurin series convergence is fastest near the point of development and decreases with distance. However, BCs are only meaningful in the context of a PA, and so the presence of the BCC in (1) cannot be explained as a property of the Maclaurin series, but as a property of the PA. The next section provides numerical support for the equivalence relationship and provides an argument for the CF dependence on the BCC.

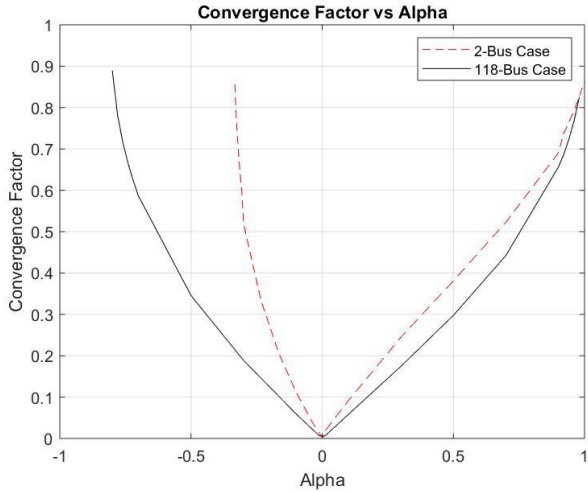

Fig. 4 CF versus α for two-bus and 118-bus systems

*B. Two-Bus Asymmetrical System Equivalence Relationship*

Recall that the equivalence of the equilibrium electrostatic charge measure (with corresponding R-N density) and the large-order PA pole/zero distribution was established for the IαP. Because the PA pole/zero distribution is a discretized version of the symmetrical equilibrium charge distribution (with expected higher (lower) density at the ends (middle) of the BC), the PA (effectively) is expected to simulate this symmetry.

Table 1 contains a tableau of pole (x)/zero (circle) plots of the corresponding diagonal, [M/M], PAs in the complex IαP and αP (real axis is horizontal, imaginary axis is vertical) as the number of α-series terms increases. These plots are limited to the domain bounded by $-1.0 \pm j2$, and $1.5 \pm j2$ in the αP and by $-3.2 \pm j1.2$ and $+1.2 \pm j1.2$ in the IαP. Observe the origin in the IαP column of Table 1 occurs at 1/3 of the distance from the RHS of this column with the scale shown in the last row of this table. The salient points upon close inspection of the tableau in Table 1 are:

- As the higher-order PAs are generated, poles and zeros occur in pairs, the roots crowd the ends of the BC while largely maintaining symmetry in the IαP, consistent with the predictions of electrostatic theory, though the zeros show more symmetry than the poles.
- Because of the restricted range, not all roots in IαP appear in the αP plots and vice versa.
- Because the origin is skewed in the plots, and due to symmetry of the roots about -1 in the IαP imposed by electrostatic principles, more roots accumulate in the negative half of the IαP, which then accumulate in the negative half of the αP, the plane where the function is to be evaluated.
- While the zeros are on the BC, convergence of poles toward the BC is consistent with the weak convergence of the measure of the PA roots distribution.

Table 1 Root Pattern for [*M/M*] PA, Asymmetrical BC

| [M/M] | Inverse-α Plane (IαP) | α Plane (αP) |
|---|---|---|
| [1/1] | | |
| [2/2] | | |
| [3/3] | | |
| [4/4] | | |
| [5/5] | | |
| ⋮ | ⋮ | ⋮ |
| [10/10] | | |
| | -3    -1    1 | -1    0    1 |

*C. Numerical Evidence of the Equivalence Principal*

To numerically verify the (weakly converging) equivalence between the PA root distribution and electrostatic charge distribution, a simulation (not shown) of the electric field for this progression of PAs, with each pole (zero) represented as a unit charge was conducted. This simulation showed that the electric field on the BC approaches zero (except, as expected, at the end points of the BC), indicating the PA root distribution is approaching the electrostatic equilibrium distribution, consistent with the developed theory.

Table 2 Root Pattern for [*M*/*M*] PA, IEEE 118 Bus System

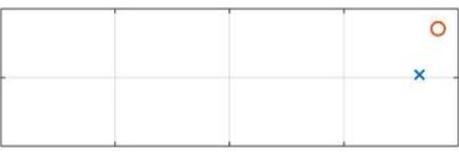

### D. The IEEE 118 bus system

The IEEE 118-bus system was solved using the canonical embedding [40]. The scale was selected so that the branch point on the positive real axis occurred at +1.0, with a corresponding BCC of 0.581. Observe that the slope of the CF vs. α curve in Fig. 4 near the origin is approximately equal to the BCC, as predicted by (1).

Table 2 contains a tableau of pole (x)/zero (circle) plots of the corresponding diagonal, [*M*/*M*], PAs in the complex IαP (real axis is horizontal, imaginary axis is vertical) as the number of α-series terms increases for the IEEE 118 bus system. These plots are limited to the domain $\pm 1.3 \pm j0.2$ in the IαP where the branch points on the real axis occur at (approx.) +1.0 and -1.248 and the scale is shown in the bottom row of the table along with root distribution of a [500/500] PA. Observe that the overall BC (implied in the bottom panel) is not terribly asymmetric and, while it is not obvious in this condense-form plot, the BC is comprised of at least 34 branch points and Chebotarev points.

This tableau demonstrates that root population in the IαP is much more complex because of the more complex topology, yet the overarching trend shows that roots tend to be distributed in a manner consistent with that of the two-bus system: the approximation of the roots to the charge distribution starts off as nearly a uniform distribution and tends toward one that has a higher concentration of roots near the branch points, which is consistent with the electrostatic charge distribution on this topology. Observe that despite the increased complexity of the BC, the CF is dictated by the number of roots needed to approximate the equilibrium distribution on the BC, which is dictated by the branch-cut's capacity and that the CF is relatively unaffected by the complexity of the topology, as expected and as predicted by (1).

## VI. CONCLUSIONS

We prove that the quantized normalized measure counting the poles and zeros of the sequence of Padé approximants converges weakly toward the continuous electrostatic charge distribution on an equivalent 2-D conducting foil and that the concept of logarithmic capacity in potential theory has the same intuitive behavior as electrostatic capacity familiar to engineers. We explain the convergence factor equation's dependence on BCC and distance from the origin. The first of these two dependencies can be explained by the convergence characteristics of the Maclaurin series. The second of these is consistent with the weak convergence of the normalized counting measure to the electrostatic charge distribution, with a larger BCC requiring more roots to approximate the equivalent equilibrium charge distribution with the same fidelity. We show that the root symmetry in the inverse-α plane predicted by the equivalence relationship between the normalized counting measure and corresponding electrostatic problem is born out in numerical experiments.